# Spatially multiplexed picosecond pulse-train generations through simultaneous intra-modal four wave mixing and inter-modal cross-phase modulation

**H. Zhang,[1,2] M. Bigot-Astruc,[3] P. Sillard,[3] and J. Fatome[1],***

[1] *Laboratoire Interdisciplinaire Carnot de Bourgogne, UMR 6303 CNRS-Université Bourgogne Franche-Comté, 9 av. A. Savary, 21078 Dijon Cedex, France*
[2] *Extreme Optoelectromechanix Laborotory, School of Physics and Electronic Sciences, East China Normal University, 500 Dongchuan Road, Minhang District, Shanghai, 200241, China*
[3] *Prysmian Group, Parc des Industries Artois Flandres, Haisnes 62092, France*

*\*Julien.Fatome@u-bourgogne.fr*

**Abstract.** We report on the experimental generation of spatially multiplexed picosecond 40-GHz pulse trains at telecommunication wavelengths by simultaneous intra-modal multiple four wave mixing and inter-modal cross-phase modulation in km-long bi-modal and 6-LP-mode graded-index few-mode fibers. More precisely, an initial beat-signal injected into the fundamental mode is first nonlinearly compressed into well-separated pulses by means of an intra-modal multiple four-wave mixing process, while several group-velocity matched continuous-wave probe signals are injected into higher-order modes in such a way to develop similar pulsed profile thanks to an inter-modal cross-phase modulation interaction. Specifically, by simultaneously exciting three higher-order modes ($LP_{11}$, $LP_{02}$ and $LP_{31}$) of a 6-LP-mode fiber along group-velocity matched wavelengths with the fundamental mode, four spatially multiplexed 40-GHz picosecond pulse-trains are generated at selective wavelengths with negligible cross-talks between all the modes.

## 1. Introduction

Stimulated by the continuously increasing data traffic in optical networks and the compelling demand to prevent a "capacity crunch" in the next decades, space-division multiplexing (SDM) in multimode fibers has emerged as a breakthrough technology for modern optical communications [1-5]. The development of SDM and associated multimode components and fibers open a new avenue for all-optical processing of spatially multiplexed signals, thus placing the nonlinear optics on the multimode platform back to the frontier of contemporary researches, among which the inter-modal nonlinear interactions in multimode fibers is an intensively investigated field of research [6-22]. On the other hand, the generation of picosecond pulse trains at high repetition rates is highly demanding in numerous applications including optical communications, optical sampling, component characterization, clock generation, metrology or spectroscopy. However, due to the limited bandwidths of usual opto-electronic components, all-optical generation technics have been proposed including active mode-locked lasers [23-25], as well as linear or nonlinear reshaping of an initial sinusoidal beat-signal into well-separated short pulses [26-38]. Among all these numerous configurations, a specific system is based on the nonlinear compression of an initial high-power beating within an optical fiber though the combined effect of Kerr nonlinearity and chromatic dispersion, namely the multiple four-wave mixing (MFWM) technique [33]. This process occurs in a single anomalous dispersive fiber and has been proved to be an attractive and efficient technique to generate high-repetition-rate pulse trains in single mode fibers [34-37] and more recently in few mode fibers (FMF) [38]. In this novel contribution, we combine the simultaneous actions of the MFWM process together with inter-modal cross-phase modulation (IXPM) [22] interactions in subsequently a bi-modal fiber and a 6-LP-mode FMF to demonstrate the generation of spatially multiplexed 40-GHz picosecond pulse trains at selective telecommunication wavelengths determined by the parameters of the FMF under test. More precisely, we first inject a high-power beat-signal into the fundamental mode as well as several continuous-wave (CW) probes into each higher-order mode supported by the FMF. We then achieve the parallel generation of well-compressed 40-GHz pulse trains in all the excited modes based on the combined actions of intra-modal multiple four-wave mixing and inter-modal cross-phase modulation. The performance of this pulse generation technique in each spatial mode has been experimentally investigated. Our experimental results are supported by the numerical simulations based on two-mode coupled nonlinear Schrödinger equations (NLS).

## 2. Principle of operation and numerical modelling

The principle of operation is schematically described in Fig. 1(a). A high-power dual-frequency pump beam ($P_1$ and $P_2$) is first injected into the fundamental mode of an anomalous dispersive few-mode fiber. In parallel, weak CW probe signals ($S_i$) are also injected into the higher-order modes supported by the fiber undertest. The crucial point here is to select the different wavelengths injected into each mode of the FMF in such a way to group-velocity match all the possible inter-modal nonlinear interactions. Basically, the different spatially multiplexed signals have to propagate with the same group-velocity so as to maximize the IXPM effect providing by the fundamental mode, while avoiding the deleterious effect of walk-off in between the generated pulses. In the frequency domain, by means of the combined effects of nonlinearity and chromatic dispersion, a Kerr frequency comb is then generated within the fundamental mode due to the MFWM process, which is concurrently spatially converted into high-order modes through IXPM onto the CW probes. In the time domain, the initial sinusoidal beating propagating into the fundamental mode is then strongly reshaped into well-separated pulses, while well-localized pulses are subsequently generated onto the temporal profile of higher-order mode probe waves at the repetition-rate defined by the frequency separation between $P_1$ and $P_2$.

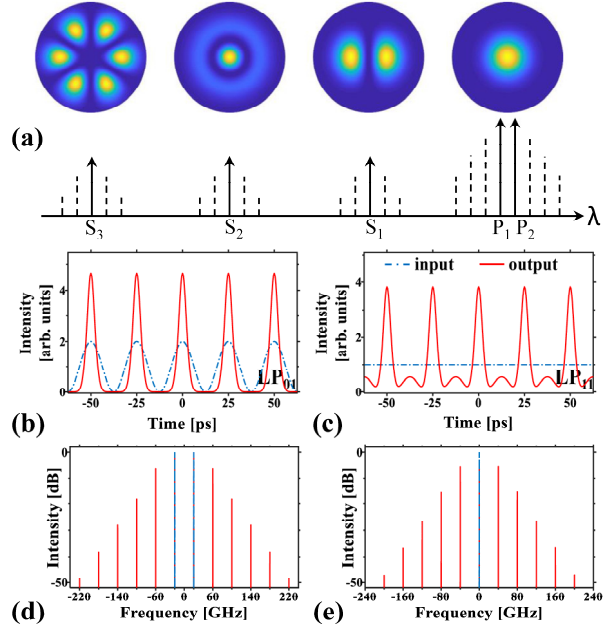

**Fig. 1.** (a) Principle of operation. A high-power dual-frequency beat-signal ($P_1$ and $P_2$) is injected into the $LP_{01}$ mode, and a weak cw signal $S_i$ is injected into each higher-order mode supported by the fiber under test. (b) The simulated temporal and spectral profiles at the output of the fiber, for the (b, d) $LP_{01}$ and (c, e) $LP_{11}$ modes. The blue dash-dotted lines represent the input profiles, and the red solid lines represent the output profiles.

To get a better description of both the intra-modal MFWM and the inter-modal XPM effects in the few-mode fiber, we use a system of two-mode coupled NLS to simulate the nonlinear interactions between a high-power beat-signal propagating in the fundamental mode and a weak CW probe injected into a higher-order mode of the FMF. The simplified scalar NLS neglecting the higher-order (third-order and higher) dispersion terms and also the Raman effect is employed as [12]:

$$\frac{\partial E_p}{\partial z} = \left[\frac{\delta_{pq}}{2}\frac{\partial}{\partial t} - \frac{i}{2}\beta_{2p}\frac{\partial^2}{\partial t^2} + i\frac{n_2\omega}{c}\left(f_{pp}|E_p|^2 + 2f_{pq}|E_q|^2\right)\right]E_p, \quad (1)$$

$$\frac{\partial E_q}{\partial z} = \left[-\frac{\delta_{pq}}{2}\frac{\partial}{\partial t} - \frac{i}{2}\beta_{2q}\frac{\partial^2}{\partial t^2} + i\frac{n_2\omega}{c}\left(f_{qq}|E_q|^2 + 2f_{pq}|E_p|^2\right)\right]E_q, \quad (2)$$

where the reference frequency $\omega$ is set at the average wavelength of the two modes, $n_2=2.6\times10^{-20}$ m$^2$/W is the silica Kerr nonlinear refractive index, $\delta_{pq} = \beta_{1p} - \beta_{1q}$ is the group velocity mismatch with $\beta_{1i}$ the inverse of the group velocity in the mode ($i=p, q$), $\beta_{2i}$ is the second-order dispersion coefficient. $1/f_{pp}$ and $1/f_{qq}$ are the effective areas of the two interacting modes with $1/f_{pq}$ the effective coupling area between them. The input conditions for each mode can be expressed as:

$$E_p(z=0,t) = \sqrt{P_p/2}\exp(-i\Delta\omega_p t), \quad (3)$$

$$E_q(z=0,t) = \sqrt{P_q/2}\exp(-i\Delta\omega_q t)\times\left[\exp(-i\Omega t/2) + \exp(+i\Omega t/2)\right], \quad (4)$$

with $\Delta\omega_i = \omega_i - \omega$ the frequency-detuning from the reference frequency, $\Omega$ the beat-frequency in units of rad/s and $P_i$ the injected power in each mode.

Note that the IXPM term in Eq. (2), which represents the action of the probe wave onto the high-power fundamental-mode beating can be neglected if weak power CWs are involved in higher-order modes.

In order to illustrate the process, we first simulate the case of a bimodal FMF. The simulated output temporal profiles of the 40-GHz beat-signal propagating in the fundamental mode and the CW probe transmitted in LP$_{11}$ are shown with red solid lines in Figs. 1(b) and 1(c), respectively. For comparison, the initial temporal profiles injected into both modes are also denoted with blue dash-dotted lines in each corresponding panel. In this simulation, the fiber parameters correspond to the two-mode graded-index fiber used in the following experiments (see Table 1). Moreover, the wavelength of the CW probe is selected to be phase-matched with the central wavelength of the beat-signal, e.g. the group velocities at the two wavelengths in each mode are equal. It can be clearly seen from Fig. 1(b) that the 40-GHz beating initially injected into the fundamental mode is then compressed into a train of well-separated picosecond pulses owing to the intra-modal MFWM. The output temporal duration is then reduced to 5 ps and negligible residual pedestals can be observed. On the other hand, a strong reshaping of the LP$_{11}$ CW probe can be observed in Fig. 1(c) with the emergence of compressed intensity spikes at a repetition rate of 40-GHz resting onto the CW background. Indeed, there still remains small bumps between adjacent peaks in the temporal profile of LP$_{11}$, due to the unavoidable interference with the residual CW background. Note that these deleterious side-lobes could be removed thanks to a nonlinear reshaping method in order to achieve a clean output intensity profile in all the modes [39-43]. In the spectral domain, here shown in Fig. 1(d) and 1(e), we can first observe the generation of a 40-GHz frequency comb in the fundamental mode due to the cascading of intra-modal FWM process (see Fig. 1(d)). Then, we also notice a clear signature of simultaneous spatial and wavelength conversion of this 40-GHz frequency comb from LP$_{01}$ to LP$_{11}$ by means of the group-velocity matched inter-modal XPM effect.

## 3. Experimental setup

The experimental setup is depicted in Fig. 2(a). Four tunable external cavity lasers (ECL) emitting in the C-band are used for this experiment. One continuous-wave ECL is first phase-modulated around 100 MHz in order to prevent any Brillouin backscattering in the fiber undertest. A $LiNbO_3$ intensity modulator (IM), driven around its zero-transmission point by a 20-GHz external RF clock is then inserted to generate the initial 40-GHz beating [32]. A programmable liquid-crystal-based optical filter (waveshaper) is further used to apply two narrow band-pass filters centered on the dual-frequency of the beating to remove undesirable residual spectral components located at 20 GHz. Afterwards, the initial beating is amplified by a single-mode Erbium doped fiber amplifier (EDFA) and then coupled into the fundamental mode together with the three other ECLs into different spatial modes of the few-mode fiber by means of a 10-mode spatial multiplexer based on a multi-plane light conversion technology [44]. Polarization controllers are also inserted into each optical path in such a way to maximize the excitation efficiency for each spatial mode. Indeed, due to random mode coupling in the graded-index FMFs undertest as well as polarization dependent performance of the spatial multiplexer and demultiplexer, we adjusted the input polarization-state of each spatial channel in such a way to optimize the energy and polarization injected into a specific degenerated mode, thus maximizing the efficiency of the inter-modal XPM process. At the output of the system, another polarization controller and fiber mechanical stress are applied on the FMF in order to maximize the energy at a specific output port of the spatial demultiplexer.

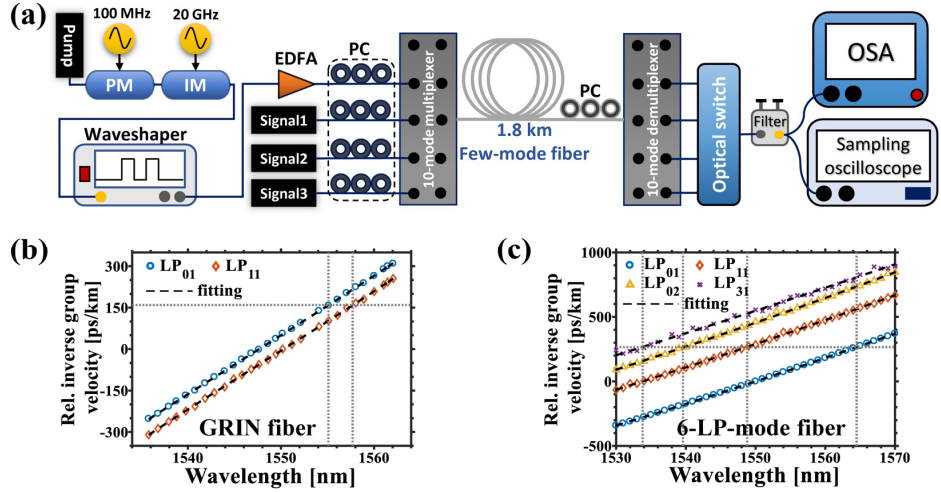

**Fig. 2.** (a) Experimental setup. PM phase modulator, IM intensity modulator, PC polarization controller, EDFA Erbium-doped fiber amplifier, OSA optical spectrum analyzer. (b-c) Experimental measurement of the relative inverse group velocity (RIGV) vs wavelength for each mode of the few-mode fibers undertest (b) Bimodal graded-index fiber (c) 6-LP-mode fiber [12]. The black dashed lines denote the linear fitting of each RIGV vs wavelength curve. The pump and signal wavelengths employed in this work are labelled by the dotted grey lines in (b) and (c).

Two FMFs are employed in this work: a bimodal graded-index fiber from *ofs* (3 spatial modes including degenerate modes) with a core diameter of 16 μm, and a 6-LP-mode graded-index fiber (10 spatial modes including degenerate modes) manufactured by *Prysmian group* with a core diameter of 22.5 μm [45]. Both fiber lengths are set to 1.8 km in order to optimize the nonlinear compression of initial beating within the fundamental mode following the design rules established in ref. [34]. The effective areas ($A_{eff}$) of each mode around 1550 nm for both fibers are listed in Table 1. The dispersion curves of each spatial mode have been measured by the time-of-flight method and shown in Fig. 2(b) and Fig. 2(c) for the bimodal fiber and 6-LP-mode fiber, respectively [12]. The measured chromatic dispersion ($D$) together with the calculated effective coupling areas ($A_{cross}$) between the fundamental mode and each higher-order mode, are also listed in Table 1. The losses for all the modes are below 0.25 dB/km, and the maximum differential-mode-group-delay (DMGD) between the spatial modes is around 55 ps/km for the bimodal fiber and lower than 550 ps/km for the 6-LP-mode fiber. Moreover, for temporal characterizations, the injection and demultiplexing conditions were optimized for each mode undertest thanks to input and output polarization controllers as well as fiber stress applied before demultiplexing operation. The total losses of the system were measured at an average value of 9.3 dB for all the modes while the average cross-talks between the different groups of modes were found better than 20 dB. Note finally, that at the output of the system, a tunable filter is also used to remove the leakage from other spatial modes.

|  | Bimodal fiber | | 6-LP-mode fiber | | | |
| --- | --- | --- | --- | --- | --- | --- |
|  | $LP_{01}$ | $LP_{11}$ | $LP_{01}$ | $LP_{11}$ | $LP_{02}$ | $LP_{31}$ |
| $A_{eff}$ (µm$^2$) | 96 | 128 | 75 | 100 | 160 | 170 |
| D (ps/km/nm) | 19.9 | 20 | 18 | 18 | 19 | 17.5 |
| $A_{cross}$ (µm$^2$) | -- | 192 | -- | 159 | 159 | 636 |

**Table 1.** Effective mode areas ($A_{eff}$), second-order chromatic dispersion (D), and effective coupling areas ($A_{cross}$) between $LP_{01}$ and each higher-order mode for both FMFs involved in the following experiments. All the values are given at 1550 nm.

## 4. Experimental results

### 4.1 Bimodal graded-index fiber

First-of-all, a 40-GHz beating with a central wavelength of $\lambda_P$=1555.1 nm is injected into the fundamental mode of the bimodal fiber, together with a CW probe signal injected into the $LP_{11}$ mode and centered on $\lambda_S$=1557.8 nm. The two wavelengths are selected owing to Fig. 2(b) to be group-velocity matched in each respective mode. The output spectra and temporal profiles of both modes have been measured as a function of the average power of the 40-GHz beating injected into $LP_{01}$, while the CW probe power is fixed to 10 dBm (all the powers are given before the spatial multiplexer input). The resulting experimental data are normalized to unity for each input power and shown in Figs. 3(a)-(d). It can be clearly seen from Fig. 3(a) that cascading intra-modal four-wave mixing sidebands are generated in $LP_{01}$ with increasing power, while its corresponding temporal profile shown in Fig. 3(b) is gradually compressed. Meanwhile, Fig. 3(c) demonstrates that the CW probe injected into the $LP_{11}$ mode also develops multiple 40-GHz sidebands due to the inter-modal cross-phase modulation process imposed by the pump beating copropagating in $LP_{01}$. Moreover, strong temporal spikes are gradually 'grown up' from the initial continuous background as shown in Fig. 3(d). The 25-ps temporal period of these generated pulses well corresponds to the repetition rate imposed by the $LP_{01}$ pump beating, thus clearly demonstrating the strong impact of this group-velocity matched inter-modal nonlinear interaction. The output pulse durations of both the pump beating in $LP_{01}$ and the signal wave in $LP_{11}$ are then reported in Fig. 3(e) as a function of the effective nonlinearity length (defined here as $L/L_{nl} = \gamma P L$ with $P$ the average power of the beating and $\gamma$ the nonlinear Kerr coefficient of the fundamental mode). We can observe that similar performance with a minimum temporal duration of 5.0 ps (full-width-half-maximum), corresponding to a duty-cycle of 1:5, is achieved in the $LP_{01}$ and $LP_{11}$ modes. In Fig. 3(f), we have compared the output temporal profiles of both modes at the optimal compression point.

We note that the initial beating is well compressed under the combined action of Kerr nonlinearity and chromatic dispersion, while similar pulsed profiles resting on the residual CW background are observed in the output signal of $LP_{11}$. A weak asymmetry can be also observed in the output intensity profile of $LP_{11}$ and is attributed to the degeneracy of this mode and associated random mode coupling between $LP_{11a}$ and $LP_{11b}$ which are characterized by slightly different group velocities.

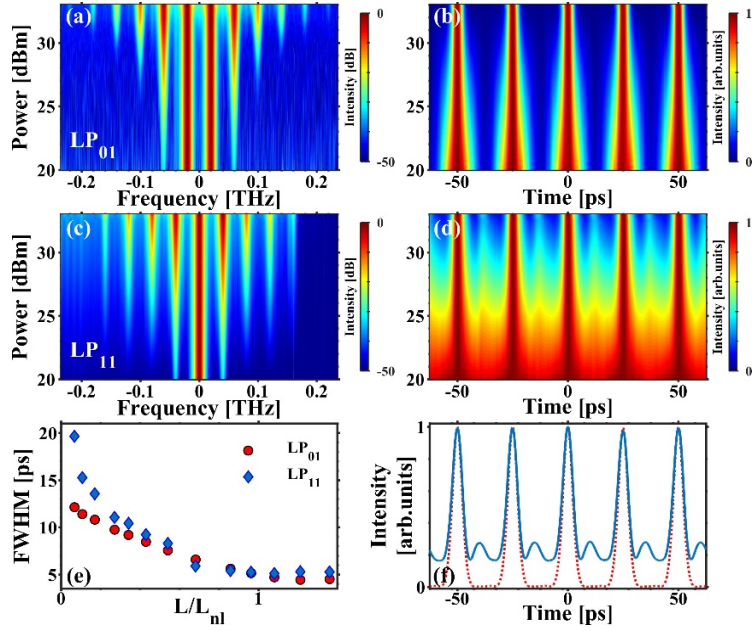

**FIG. 3.** Spectral and temporal evolution of the output pulse-trains in the bimodal FMF. The generated pulse-trains in the fundamental mode $LP_{01}$ (a, b) and the higher-order mode $LP_{11}$ (c, d) are measured as a function of the injected power of the 40-GHz beating. (e) Output pulse duration (full-width-half-maximum) as a function of the effective nonlinearity length $L/L_{nl}$ calculated for the fundamental mode (see text). (f) Intensity profiles of $LP_{01}$ (red dotted line) and $LP_{11}$ (blue solid line), recorded at the optimal compression point, are shown in.

Next we investigate the dependence of the output signal generated in $LP_{11}$ on the wavelength separation between the two modes ($\Delta\lambda=\lambda_S-\lambda_P$), that is to say the impact of a group-velocity mismatch or walk-off effect. The wavelength and power of the $LP_{01}$ beating is fixed at 1555.1 nm and 32 dBm, corresponding to the optimal compression point, while the wavelength separation of the 10-dBm $LP_{11}$ CW probe is scanned from 2.2 to 3.2 nm. It is clearly observed from Fig. 4(a) and Fig. 4(b) that asymmetric spectral and temporal profiles are generated as the signal wavelength moves away from the phase-matched wavelength (represented by the white dashed lines in both panels). Two examples of output spectra are plotted in Fig. 4(c), for which the blue-shifted sidebands surpass the red-shifted sidebands as the signal wavelength is red-detuned ($\Delta\lambda=3.1$nm, red dashed lines), while the red-shifted sidebands beat the blue-shifted sidebands for an opposite blue-shifted signal ($\Delta\lambda=2.3$ nm, blue solid lines). Meanwhile, the temporal profiles of the output signal for the two cases are plotted in Fig. 4(d), with a dramatical temporal broadening and tilting compared to the optimally compressed signal shown in Fig. 3(f), thus highlighting the strong sensitivity and impact of the group-velocity matching condition.

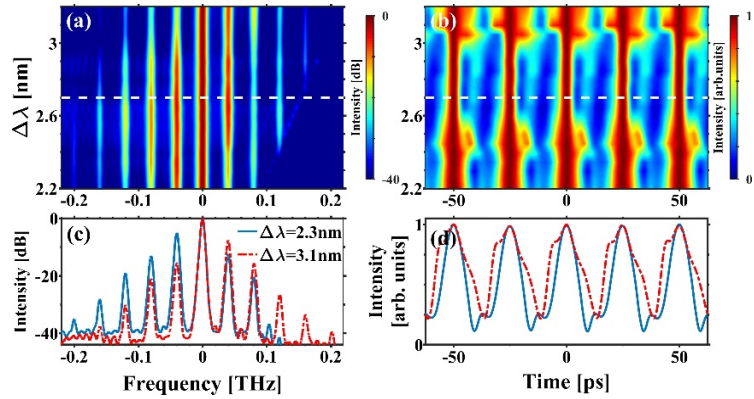

**Fig. 4.** Spectral (a) and temporal (b) profiles of the output $LP_{11}$ CW probe with respect to the wavelength separation $\Delta\lambda$ from the central wavelength of the $LP_{01}$ beating. Two typical spectra and temporal profiles are plotted in (c) and (d), corresponding to $\Delta\lambda=2.3$ nm (blue solid line) and $\Delta\lambda=3.1$ nm (red dash-dotted line).

*4.2 6-LP-mode fiber*

Similar measurements are further conducted in the *Prysmian group* 6-LP-mode fiber. The central wavelength of the initial beating is changed to $\lambda_P=1565$ nm, in order to accommodate the phase-matched wavelengths of all higher-order modes within the C-band, considering the larger DMGD of this FMF. Note that the group-velocity matching in this 6-LP-FMF is then achieved for negative values of wavelength separation, in contrast to the previous bimodal fiber. A CW signal centered at 1549 nm is first injected into the first higher-order mode $LP_{11}$, and the output spectra and temporal profiles of $LP_{01}$ and $LP_{11}$ are measured as a function of the input average power of the $LP_{01}$ beating. As previously reported in the bimodal fiber, it can be clearly seen from Figs. 5(a)-(d) that a 40-GHz frequency comb is gradually generated in both modes with increasing power, and the corresponding temporal profiles are compressed due to the combined effects of Kerr nonlinearity and anomalous chromatic dispersion. The pulse durations of the output temporal profiles for both modes are plotted in Fig. 5(e) as the function of the effective nonlinearity length in the fundamental mode. Similar behaviors are identified for $LP_{01}$ and $LP_{11}$, with an optimally compressed duration of 4.9 ps corresponding to a duty-cycle of 1:5. Besides, the output temporal profiles of $LP_{01}$ and $LP_{11}$ at the optimal compression point are depicted in Fig. 5(f), where a pedestal-free picosecond pulse train is generated in $LP_{01}$ while similar pulsed profiles are induced in $LP_{11}$ though with a low continuous-wave background. These results remain quite similar to the previous case obtained in the bimodal FMF.

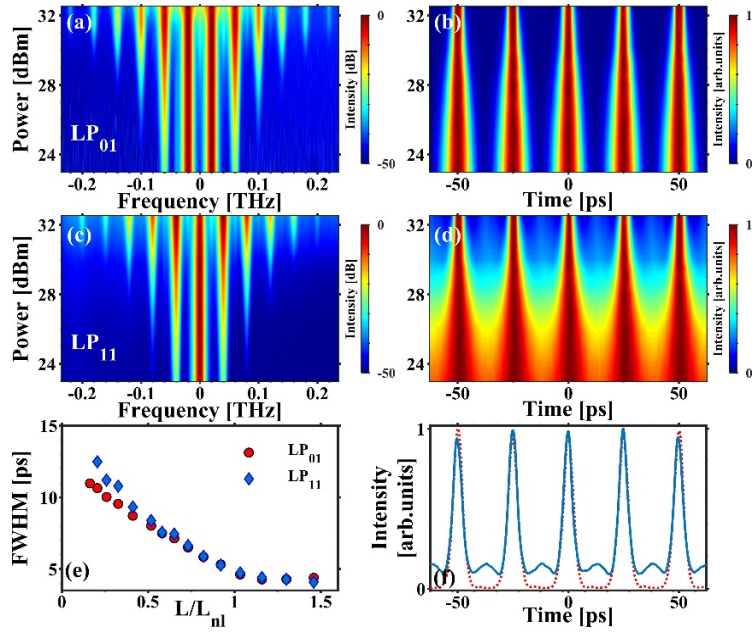

**FIG. 5.** Spectral and temporal evolution of the output pulse-trains in the 6-LP-mode fiber for $LP_{01}$ and $LP_{11}$. Figure caption is same as in Fig. 3.

The dependence of the output signal generated in $LP_{11}$ on the wavelength separation ($\Delta\lambda=\lambda_S-\lambda_P$) between the two first modes of the 6-LP-mode fiber is also further investigated. The wavelength of the beating injected into $LP_{01}$ is fixed at 1565 nm with its power set at the optimal point. As can be clearly seen from Fig. 6(a) and Fig. 6(b), the output spectra and temporal profiles from $LP_{11}$ become asymmetric and distorted when the signal wavelength is shifted from the phase-matched wavelength (represented by the white dashed lines in both panels).

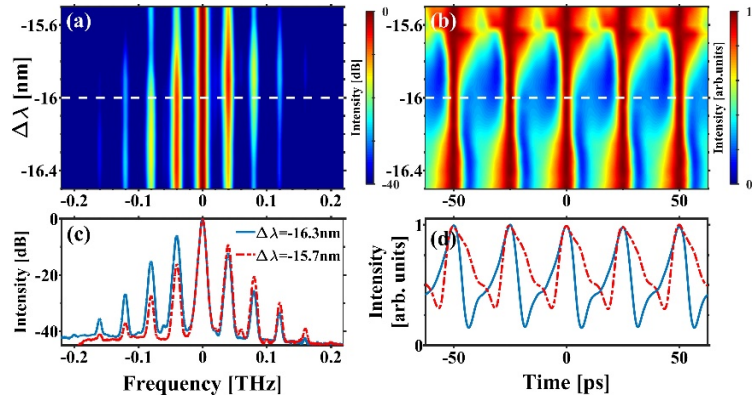

**FIG. 6.** Output spectral (a) and temporal (b) profiles of the CW $LP_{11}$ signal with respect to its wavelength separation $\Delta\lambda$ from the $LP_{01}$ central wavelength. Two typical spectra and temporal profiles are plotted in (c) and (d), corresponding to $\Delta\lambda=-16.3$ nm (blue solid line) and $\Delta\lambda=-15.7$ nm (red dash-dotted line).

In Figs. 6(c)-(d) two output spectra and corresponding temporal profiles are plotted with the signal wavelength red-shifted ($\Delta\lambda=-15.7$ nm, red dashed lines) and blue-shifted ($\Delta\lambda=-16.3$ nm, blue solid lines), respectively. The output temporal profiles are broadened and tilted when the input signal wavelength is unphase-matched with the central wavelength of the beating. Once again, these results underline the high sensitivity of the inter-modal nonlinear interactions in FMFs with respect to the phase matching conditions, as already highlighted in several previous works [7-12] and which could be seen as a strong limitation for practical applications without a specific fiber design [46-48].

### 4.3 Simultaneous inter-modal cross-phase modulation in three higher-order modes

Finally, we have studied the simultaneous inter-modal XPM interactions between the initial 40-GHz pump beating injected into the fundamental mode and three CW probe signals propagating in three higher-order modes ($LP_{11}$, $LP_{02}$, $LP_{31}$) of the 6-LP-mode FMF. The central wavelength of the 40-GHz beating injected into the fundamental mode is set to 1565 nm, while the three other signal wavelengths are adjusted to be group-velocity matched in each higher-order mode. In agreement with our measurements reported in Fig. 2(c), the optimum probe wavelengths are found to be 1549 nm ($\Delta\lambda$=-16 nm) for $LP_{11}$, 1540 nm ($\Delta\lambda$=-25 nm) for $LP_{02}$, and 1534 nm ($\Delta\lambda$=-31 nm) for $LP_{31}$.

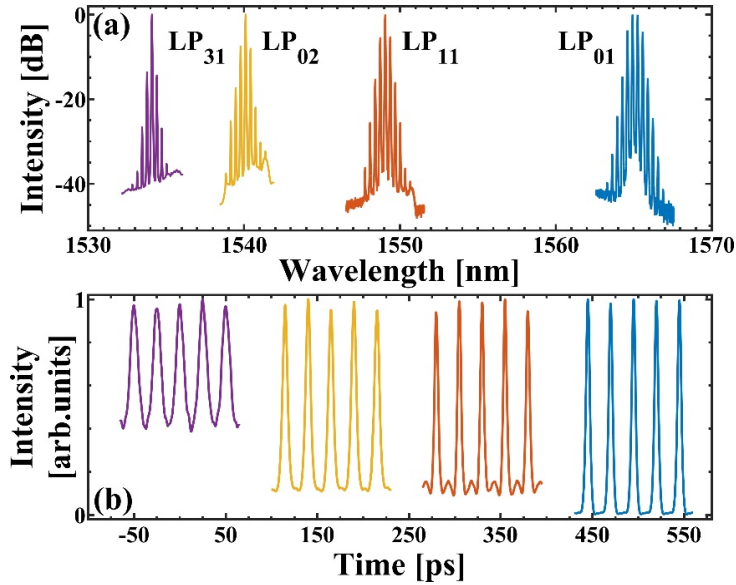

**FIG. 7.** Output spectra (a) and temporal profiles (b) recorded in $LP_{01}$ (blue), $LP_{11}$ (orange), $LP_{02}$ (yellow), and $LP_{31}$ (purple) at the output of the 6-LP-mode fiber. The temporal profiles shown in (b) are horizontally shifted for better clarity.

The output spectra and temporal profiles from each spatial mode demultiplexed at the output of the 6-LP-mode fiber are measured and depicted in Fig. 7(a) and Fig. 7(b), respectively. We can observe from Fig. 7(a) that the 40-GHz frequency comb generated in $LP_{01}$ is simultaneously converted into all the three higher-order modes through inter-modal XPM processes. We can note however that the conversion efficiency decreases for highest order modes especially for $LP_{31}$ where the conversion efficiency (here defined as the power ratio of the first sideband over the signal wavelength) is below -15 dB, while it achieves almost -5 dB for other spatial modes. This behavior can be attributed to the larger effective coupling areas between higher-order modes and the fundamental mode, as well as the random coupling occurring in between the degenerate modes belonging to the same mode-group (i.e. the third group $LP_{02}$+$LP_{21a}$+$LP_{21b}$, the fourth group $LP_{31a}$+$LP_{31b}$+$LP_{12a}$+$LP_{12b}$). Regarding the temporal intensity profile recorded at the output of the spatial demultiplexer for each higher-order mode and shown in Fig. 7(c), we can observe the generation of well-separated pulses with a symmetric shape and low residual background due to the well-maintained group-velocity matching with the fundamental mode beating. Only the highest-order mode $LP_{31}$ presents a high level of CW background and less compressed intensity profile due to the modest nonlinearity undergone by this group of modes. This behavior is attributed to the increasing number of degenerate modes in this group (4 in total) and their associated random mode couplings during propagation, for which the slight differences in the phase matching condition enlarge the resulting bandwidth at the expense of the conversion efficiency as well as a weak random walk-off between the degenerated modes. The nonlinear conversion in the $LP_{31}$ spatial mode is also less efficient due to a large mode area and thus a weak nonlinearity.

Note also that the input state-of-polarization for each spatial mode has been carefully adjusted, especially for highly degenerated modes, in order to optimize the output temporal profile at the output of our spatial demultiplexer. Nevertheless, fluctuations over time of the quality of the generated pulse train have been observed, particularly in the highest-order degenerate mode ($LP_{31}$). This behavior is attributed to the evolution of the linear mode coupling which randomly spreads the energy in the whole group of modes due to external perturbations (temperature, vibration…). However, these fluctuations occur in a time scale of several minutes and can be compensated by readjusting the whole experimental conditions (polarization controllers and FMF mechanical stress). We have checked the possible impact of the cross-talks between the different modes under study, by comparing the outputs from each higher-order mode with and without the other higher-order modes excited. No noticeable difference was observed both in the spectrum and temporal profile, due to the high quality of multiplexing/demultiplexing operation (>20 dB) of the 10-mode spatial multiplexer/demultiplexer.

## 5. Conclusion

To summarize, we have reported the simultaneous generation of spatially multiplexed picosecond 40-GHz pulse trains at telecommunication wavelengths through the synergic actions of intra- and inter-modal nonlinear interactions occurring in km-long bi-modal and 6-LP-mode graded index few-mode fibers. The principle of operation is based on the nonlinear compression of a high-power initial beating propagating in the fundamental mode into well-separated pulses by means of an intra-modal multiple four wave mixing process. Subsequently, group-velocity matched CW probes are injected into higher-order modes and undergone a strong temporal reshaping due to the combined effects of inter-modal cross-phase modulation and anomalous dispersion. In the aim to provide an experimental proof-of-principle experiment, we have successfully generated four spatially multiplexed well-separated 40-GHz picosecond pulse trains at selective telecommunication wavelengths in a 6-LP-mode graded index fiber. These experimental results have also highlighted the dramatic sensitivity of the inter-modal nonlinear interactions occurring in FMFs with respect to the phase matching conditions. Indeed, for practical applications, a careful design of the FMF would be required in order to first increase the phase-matching bandwidth, secondly to increase the nonlinearity and finally to limit the impact of random mode coupling in between degenerated modes. It is also important to notice that similar results could be obtained in a step-index few mode fiber and could be extended to a higher number of modes. Moreover, since the group velocity mismatch between higher-order modes is larger in a step-index FMF, it should be then possible to spatially convert frequency combs on a larger range of wavelengths while satisfying the group-velocity matching condition. In conclusion, these results demonstrate that spatial multiplexing of nonlinear functionalities can be performed in few-mode fibers and open the way toward the design of all-optical signal processing tools for spatial division multiplexing applications.

**Funding**

This work was founded by the Agence Nationale pour la recherche ANR APOFIS project ANR-17-ERC2-0020-01.

**Declarations of interest:** none

**Acknowledgments:** This work benefits from the PICASSO platform in ICB.